\pdfoutput=1
\documentclass[a4paper,11pt]{article}
\usepackage{pos}
\usepackage{amsmath}
\usepackage{xspace}
\renewcommand{\logo}{\relax}

\title{Vector boson production in association with jets in CMS}

\manuallySeparateAuthors
\author*[a]{Philippe Gras}
\author{on behalf of the CMS Collaboration}

\affiliation[a]{IRFU, CEA, Universit\'e Paris-Saclay\\
  Gif-sur-Yvette, France}

\emailAdd{philippe.gras@cern.ch}

\abstract{The study of the associated production of vector bosons and jets constitutes an excellent environment to check numerous QCD predictions. Total and differential cross sections of vector bosons produced in association with jets have been studied in pp collisions using CMS data. Differential distributions as a function of a broad range of kinematical observables are measured and compared with theoretical predictions. In this talk, studies of associated production of vector bosons with inclusive jets and with jets originating from heavy-flavour quarks will be summarized.}

\FullConference{%
  41st International Conference on High Energy physics - ICHEP2022\\
  6-13 July, 2022\\
  Bologna, Italy
}

\newcommand{\pt}{\ensuremath{p_{\text{T}}\xspace}}

\newcommand{\TeV}{\ensuremath{\,\text{Te\hspace{-.08em}V}}\xspace}
\newcommand{\GeV}{\ensuremath{\,\text{Ge\hspace{-.08em}V}}\xspace}
\newcommand{\MeV}{\ensuremath{\,\text{Me\hspace{-.08em}V}}\xspace}

\begin{document}

\maketitle

\section{Introduction}

Vector bosons are abundantly produced at the LHC in association with jets and the measurement of the integrated and differential cross sections of the production processes provide a good test of perturbative QCD (pQCD) calculations.  As backgrounds for many LHC analyses, the validation of the simulation of these processes and the control of the produciton cross section calculations are essential. They also allow validation of calculation techniques common with the Higgs production cross section calculations. In addition, the vector bosons produced in association with jets offer a playground to explore non-perturbative terms, like soft gluon radiations and parton distribution functions (PDFs).

The most recent results from the CMS Collaboration on the associated production of vector bosons and jets have been presented at the 41st International Conference on High Energy physics.
 
\section{Differential cross section of production of a Z boson in association with jets}

The Z+jets results presented here extend the previous CMS 13\TeV measurement from the data set collected in 2015 using a 16 times larger data set acquired during the year 2016 and representing $35.9\,\text{fb}^{-1}$. They cover numerous differential cross sections, with one or two dimensions and are described in three papers submitted for publication~\cite{CMS:2022ilp,CMS:2022ubq,CMS:2022vkb}. Comparison with Monte-Carlo generator predictions have been extended compared to previous publications with predictions  based on the parton branching (PB) method with transverse-momentum dependent (PB-TMD) parton distribution functions (PDFs)~\cite{BermudezMartinez:2018fsv} together with
a TMD-based parton showering (PS), \textsc{Cascade3} (CA3)~\cite{Baranov:2021uol}. 

\begin{figure}[p]
	\hfill
	\includegraphics[width=0.3\textwidth]{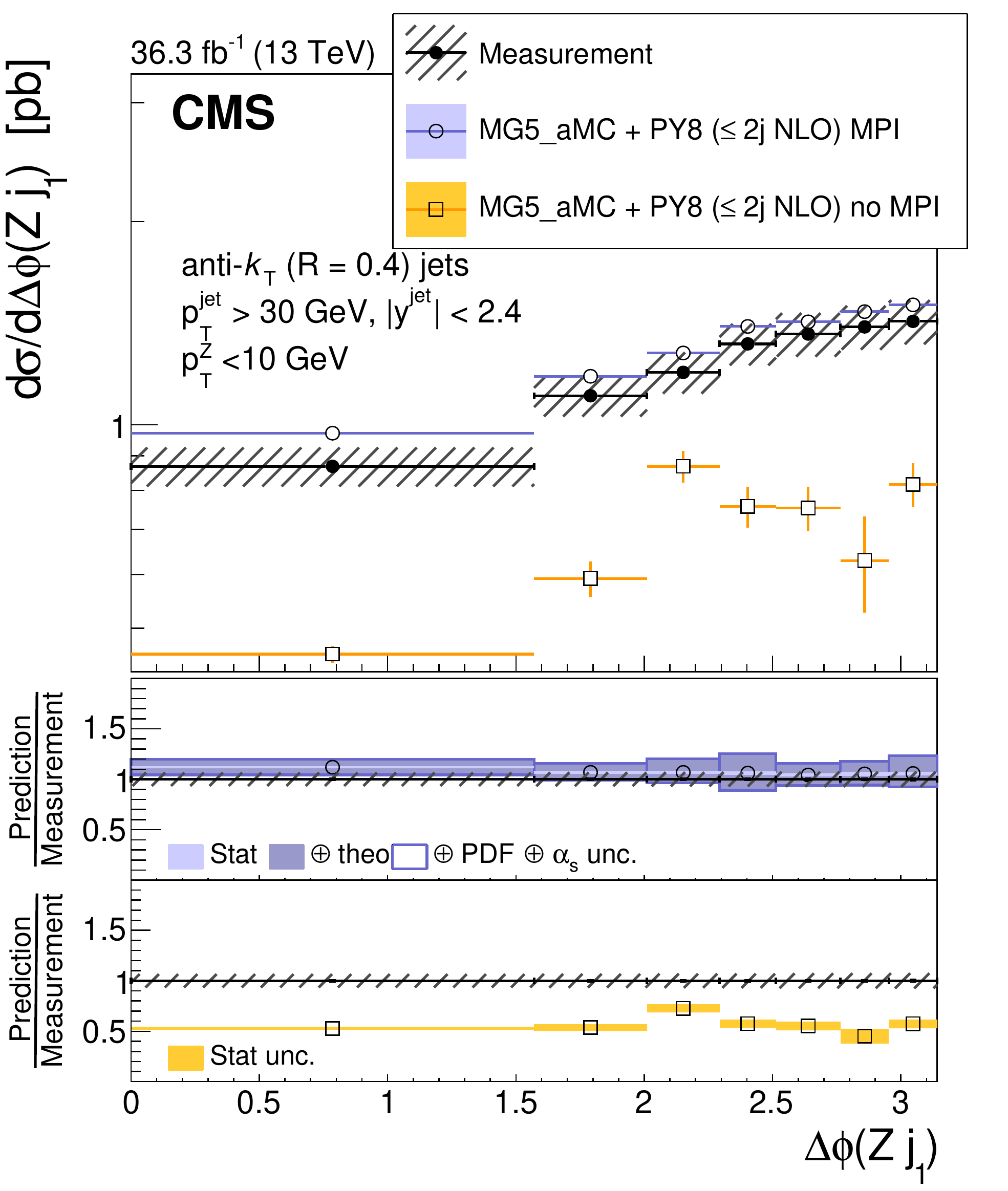}\hfill
	\includegraphics[width=0.3\textwidth]{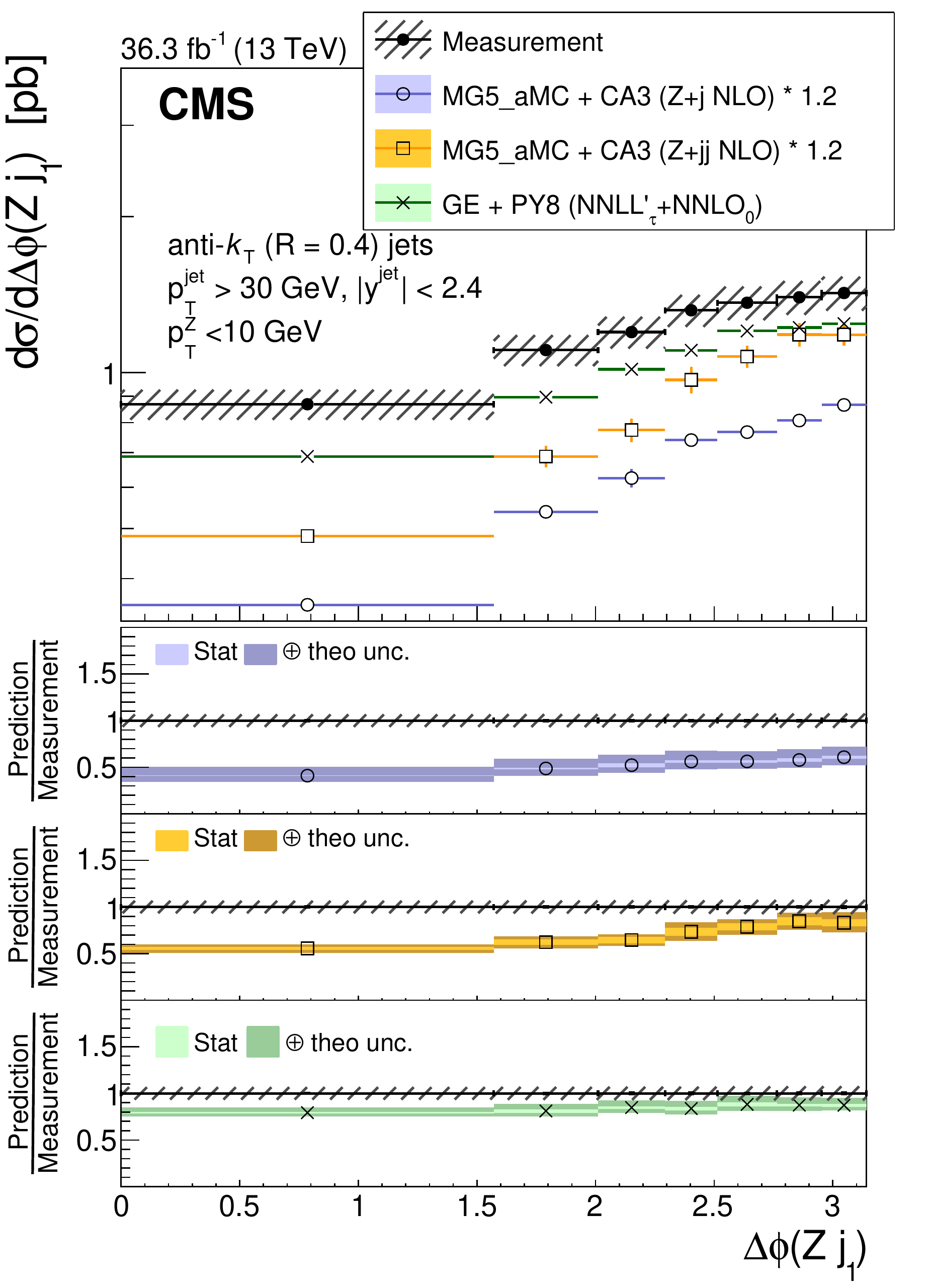}\hfill
	\includegraphics[width=0.3\textwidth]{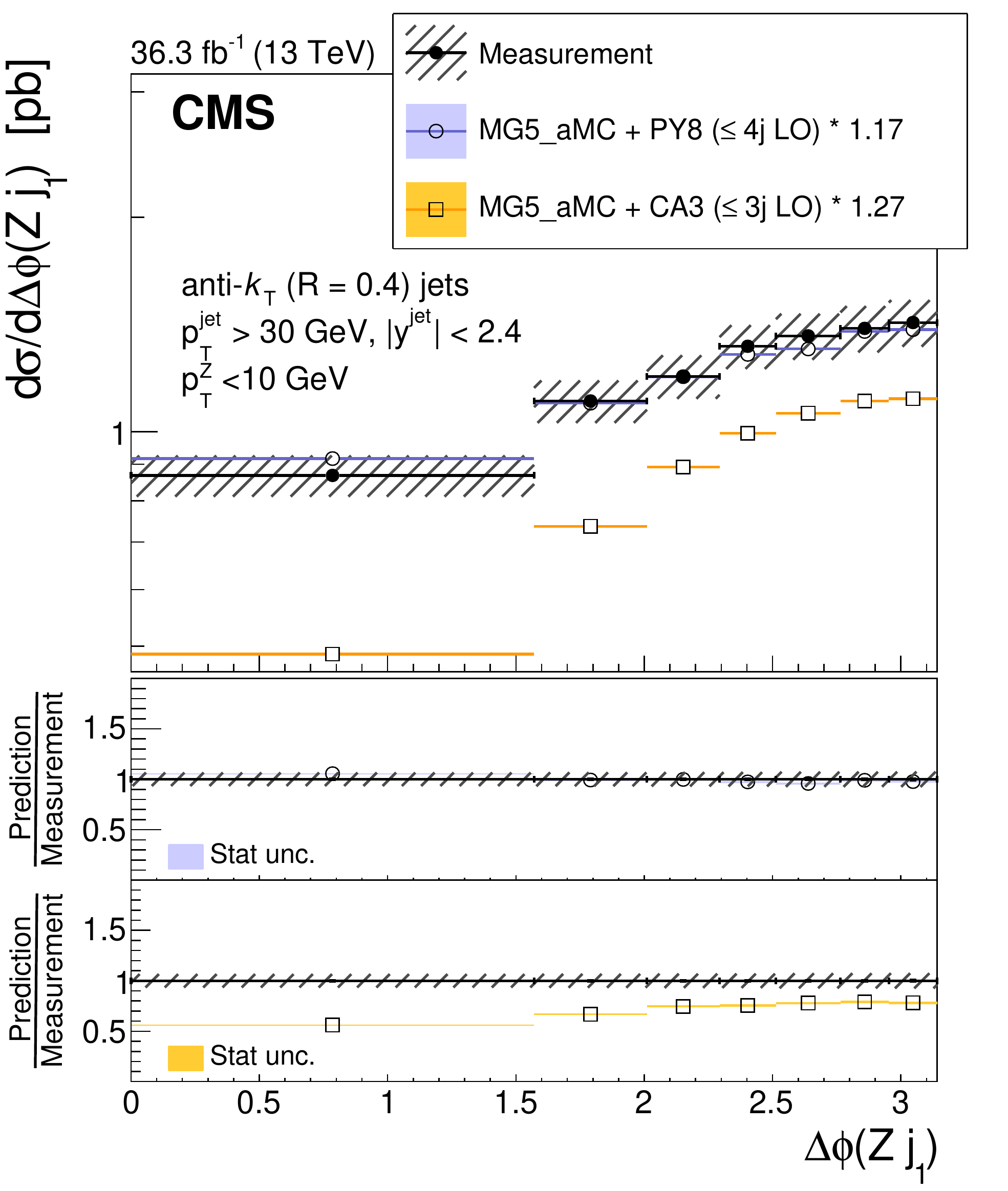}\hfill\vspace{0pt}
	\caption{Cross section as a function of the azimuthal angle between the Z boson and the leading jet, $\Delta \phi_{\text{Z},\text{j}_1}$, in the region \pt (Z) $\le$ 10\GeV and for event with at least one jet with $\pt>30\GeV$ in the rapidity acceptance~\cite{CMS:2022vkb}. The error bars on the data points represent the statistical uncertainty of the measurement, and the hatched band shows the total uncertainties of statistical and systematic sources added in quadrature.}\label{fig1}
\end{figure}

In particular, the angular correlation, in the transverse plane (azimuthal angle) and longitudinally to the beam (pseudorapidity), between the Z boson and the jets have been studied in three different regimes, a boson with $\pt$ much lower than the required jet, $\pt(\text{Z}) \le 10\GeV$, around the jet $\pt$ threshold, $\pt(\text{Z}) \in (30, 50\GeV)$, and with a large value, $\pt(\text{Z})>100\GeV$ for a dilepton in the Z mass peak region. In the region with a $\pt$ much smaller than the dilepton mass, a soft gluon radiation summation at all orders is required for an accurate cross section calculation. Good modelling of this region is essential for the W boson mass measurement. Fig.~\ref{fig1} shows the results for the azimuthal angle in the low-$\pt$ region, which is dominated by multiple parton interactions ($\approx 50\%$ as observed on the left plot) and dijet events with a Z boson radiated from a parton. The {\sc MG5\_aMC} FxFx prediction with {\sc Pythia8} and the CUETP8M1 tune~\cite{CMS:2015wcf} shows an excellent agreement with the measurement. The calculation includes matrix elements at next-to-leading order (NLO) in pQCD for final states with 0, 1 and 2 jets, merged with the FxFx scheme~\cite{Frederix:2012ps}. The slope of the ratio of {\sc Geneva} + {\sc Pythia8} to the measurement is small, indicating that the discrepancy could come from the underlying event modelling and calls for a dedicated {\sc Pythia8} tune for this generator. The {\sc MG5\_aMC} + {\sc Cascade}~3 predictions, including in the matrix elements processes with a final state with one (denoted 1j), two partons (denoted 2j) at NLO or 0, 1, 2, and 3 partons at LO (denoted $\le$ 3j LO), are failing to describe the measurement. They do not include the underlying events.

\begin{figure}[p]
  \hfill
  \includegraphics[width=0.3\textwidth]{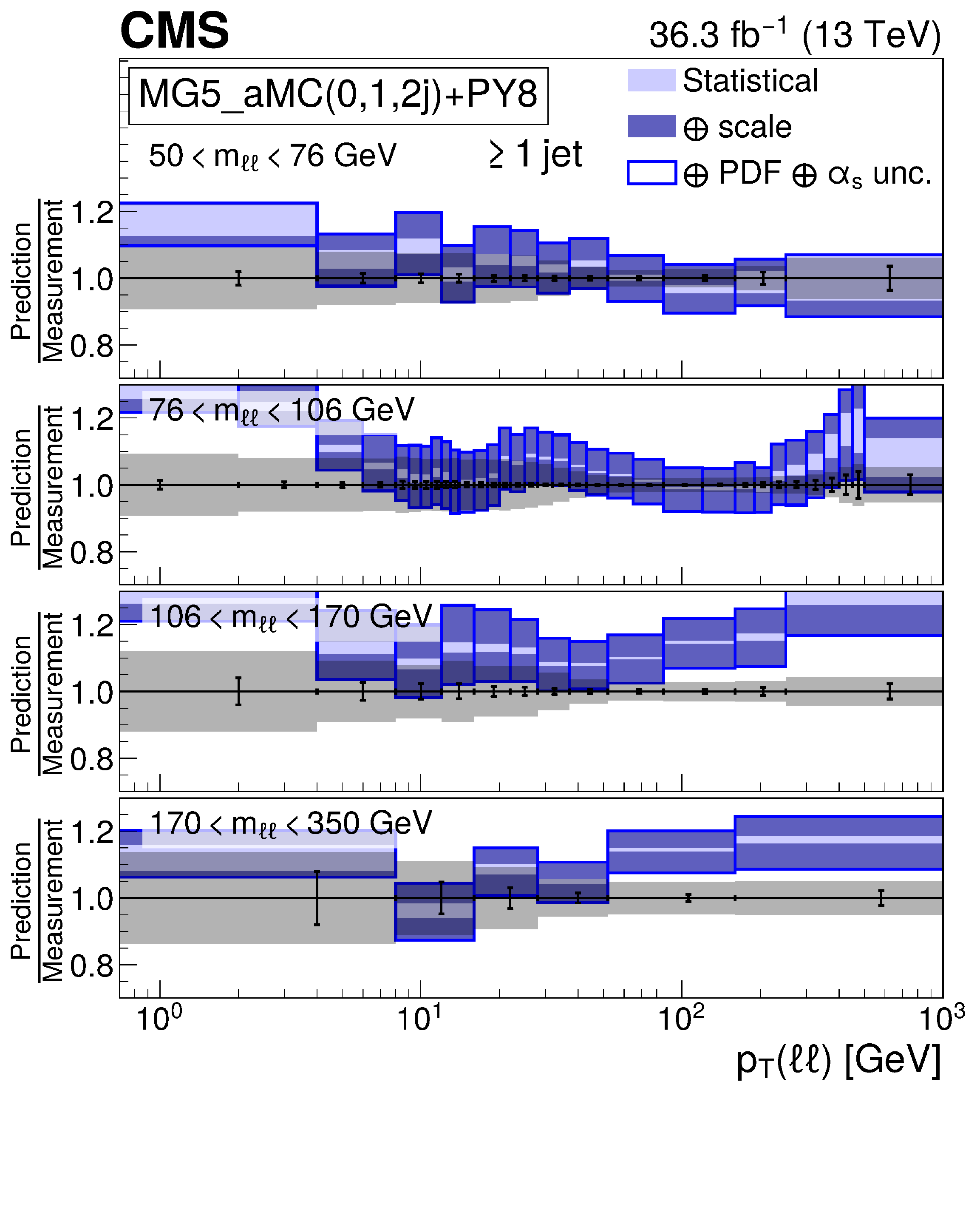}\hfill
  \includegraphics[width=0.3\textwidth]{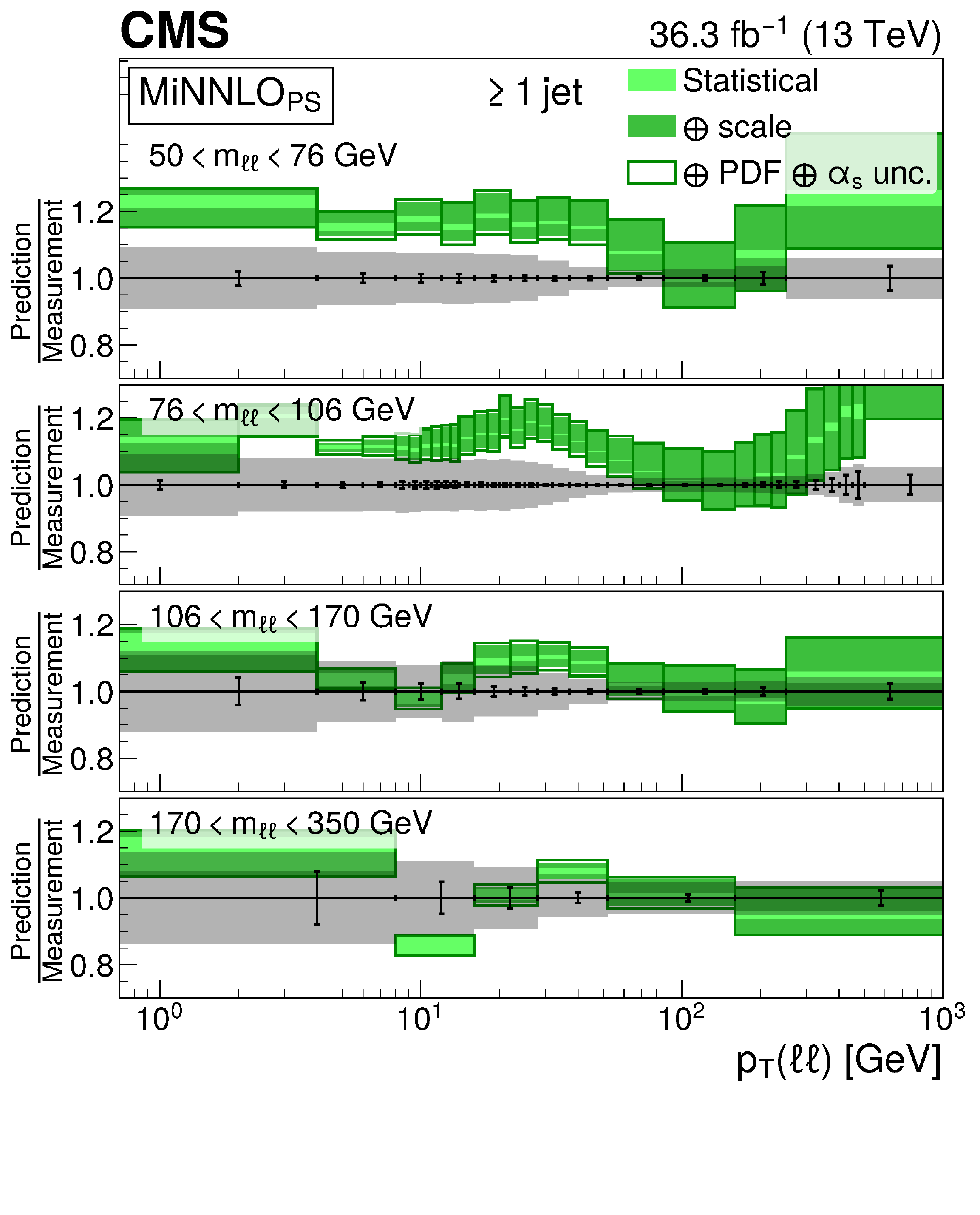}\hfill\vspace{0pt}\\
  \vspace{0pt}\hfill
  \includegraphics[width=0.3\textwidth]{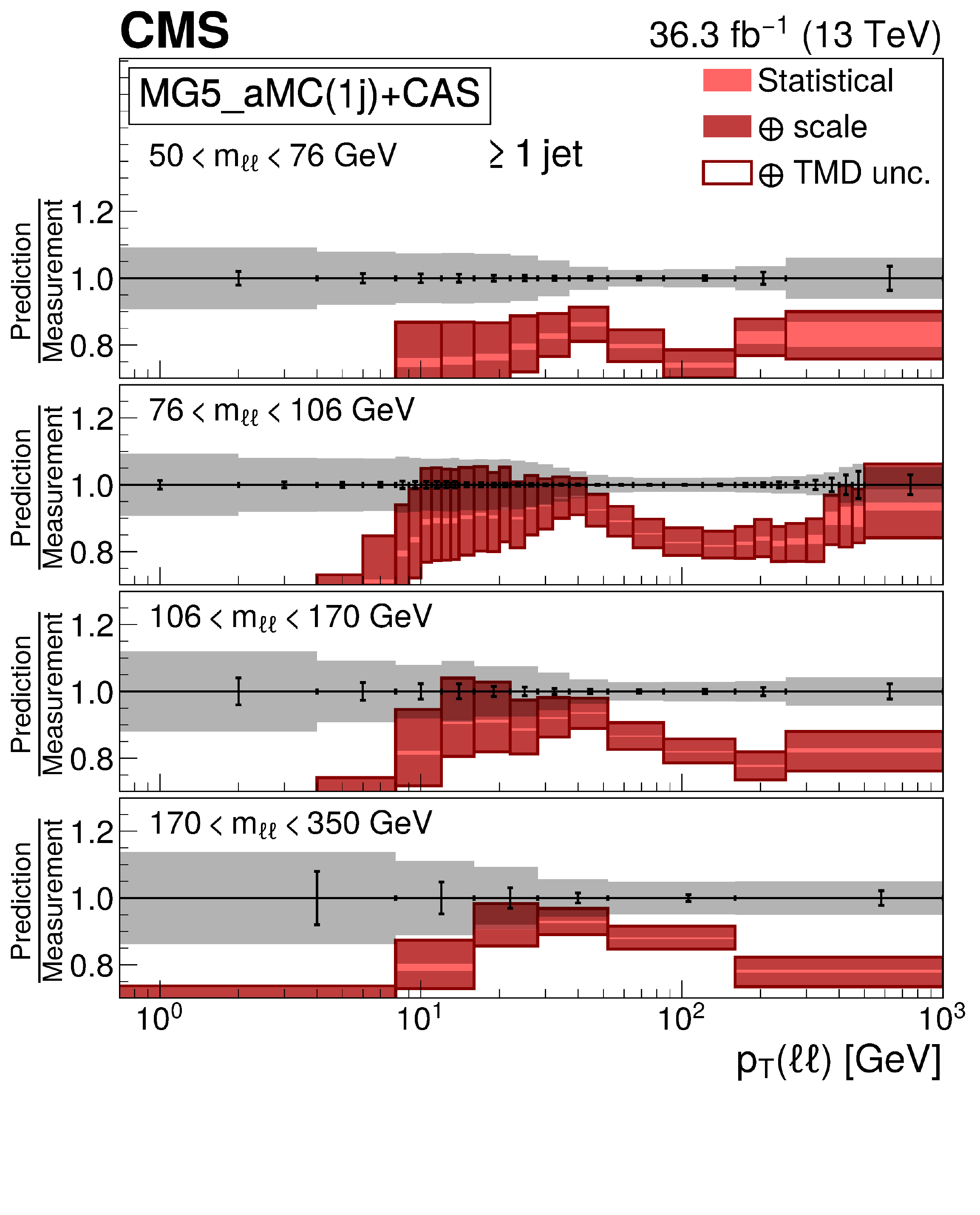}\hfill
  \includegraphics[width=0.3\textwidth]{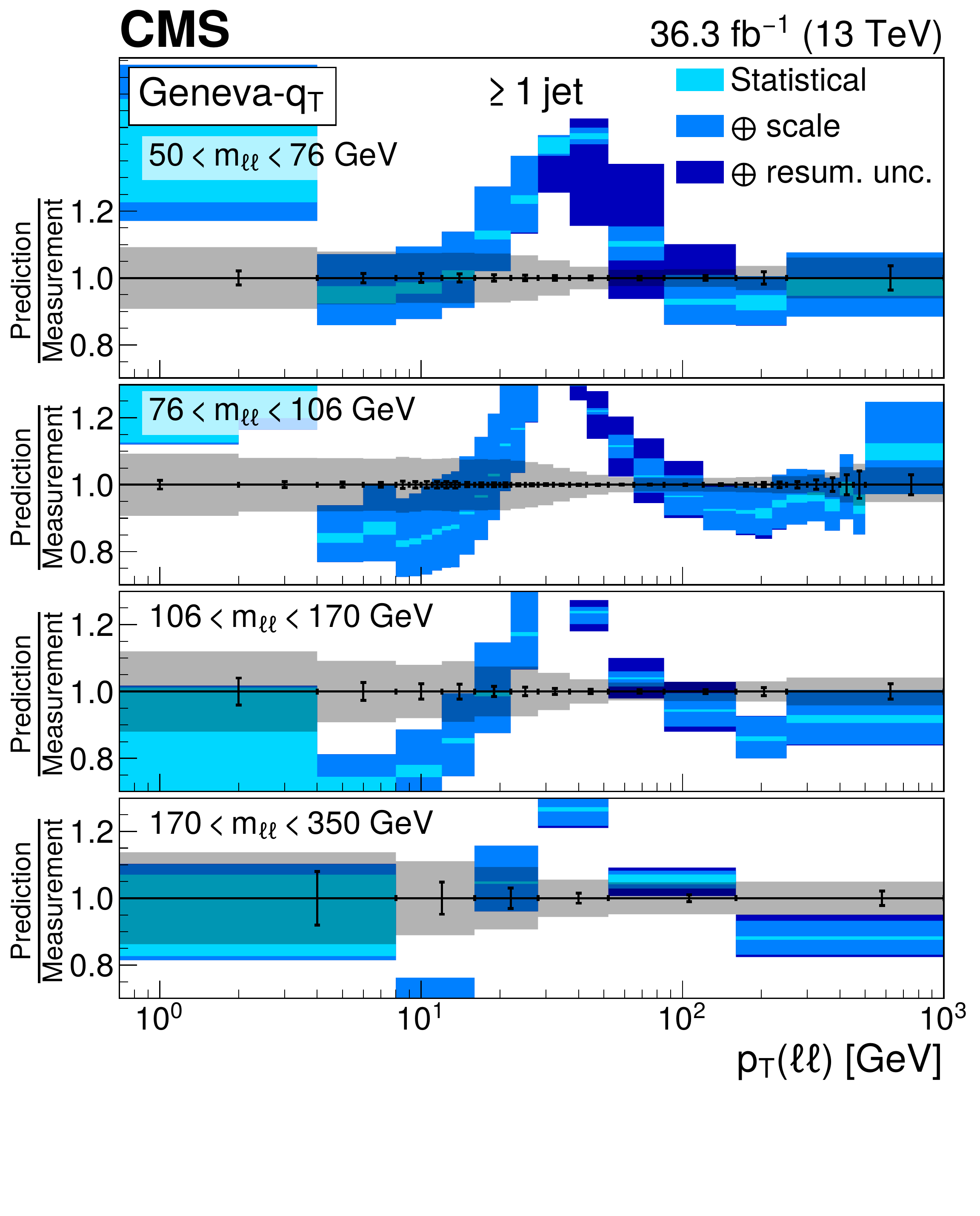}\hfill\vspace{0pt}
  \caption{Comparison of the differential cross sections in $\pt(\ell\ell)$ to different predictions in various invariant mass ranges for event with at least one $\pt>30\GeV$ jet in the rapidity acceptance~\cite{CMS:2022ubq}. The error bars on the data points represent the statistical uncertainty of the measurement, and the hatched band shows the total uncertainties of statistical and systematic sources added in quadrature.}
  \label{fig2}
\end{figure}

\section{Mass dependency of dilepton pair on \texorpdfstring{p$_{\text{T}}$}{p\_T} in Drell-Yan production}

The dilepton $\pt$ distribution in the Drell-Yan process has also been studied on a wide dilepton mass range from 50\GeV to 1\TeV extending the measurements restricted to the Z boson peak. The differential cross section is measured as functions of $\pt(\text{Z})$ and $m(\text{Z})$~\cite{CMS:2022ubq}.

The measurement for events with at least one jet with $\pt$ above 30\GeV in the $|y|<2.5$ acceptance is compared to different predictions in Fig.~\ref{fig2}. In addition to the predictions described before, we have compared this measurement with a Monte-Carlo sample obtained with {\sc MiNNLO$_{\text{PS}}$}~\cite{Monni:2019whf,Monni:2020nks} and {\sc Pythia8}.

The region with $\pt(\ell\ell)$ below the jet \pt threshold requires either another jet to balance the event or multiple parton interactions. The samples from {\sc MG5\_aMC} FxFx + {\sc Pythia8} and {\sc MiNNLO$_{\text{PS}}$} provide the best description of the data, although none of the predictions can describe them perfectly in the full mass range.

\section{Production in association with heavy flavor jets}

\begin{figure}[htpb]
	\hfill
    \includegraphics[width=0.34\textwidth]{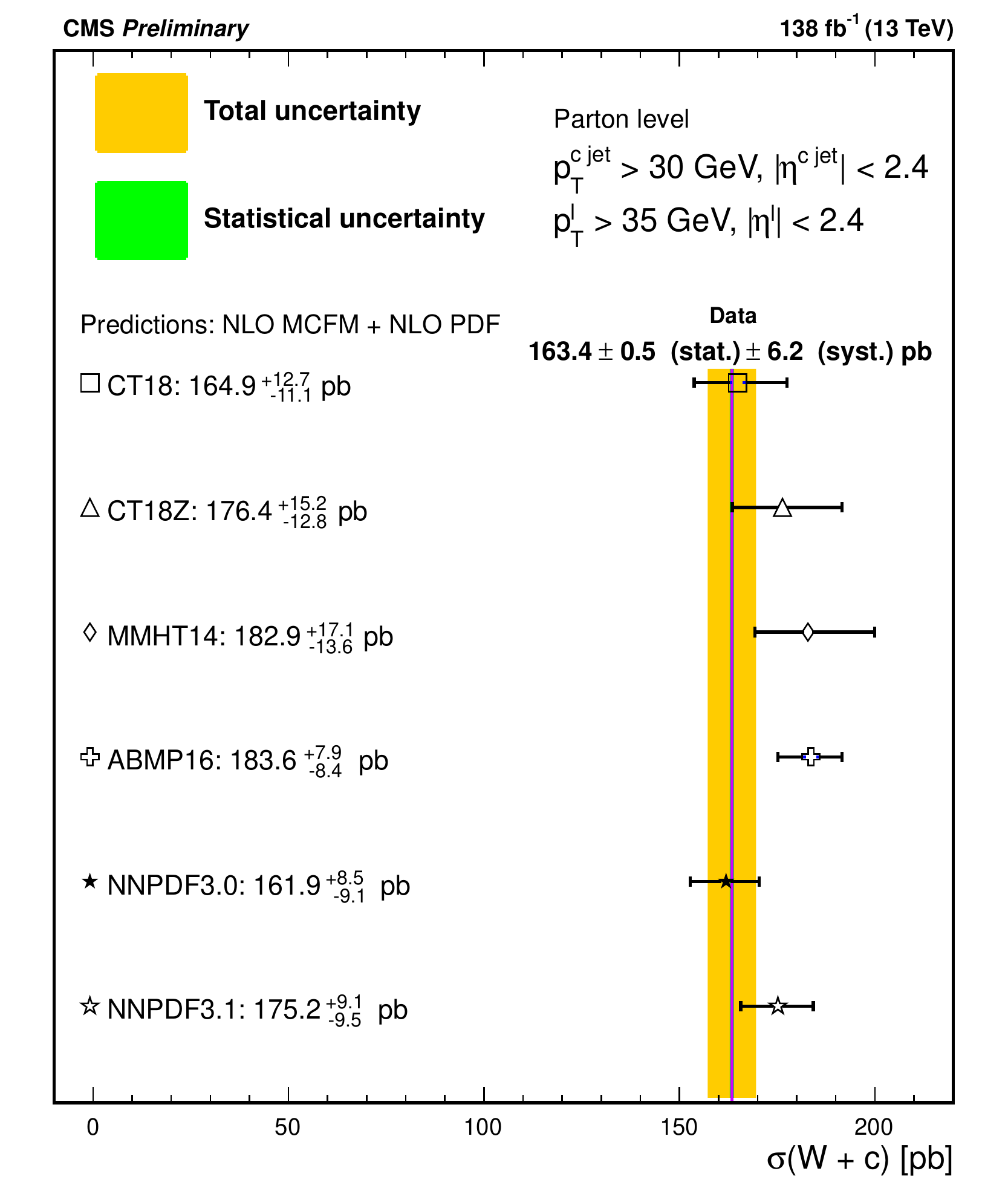}\hfill
		\includegraphics[width=0.34\textwidth]{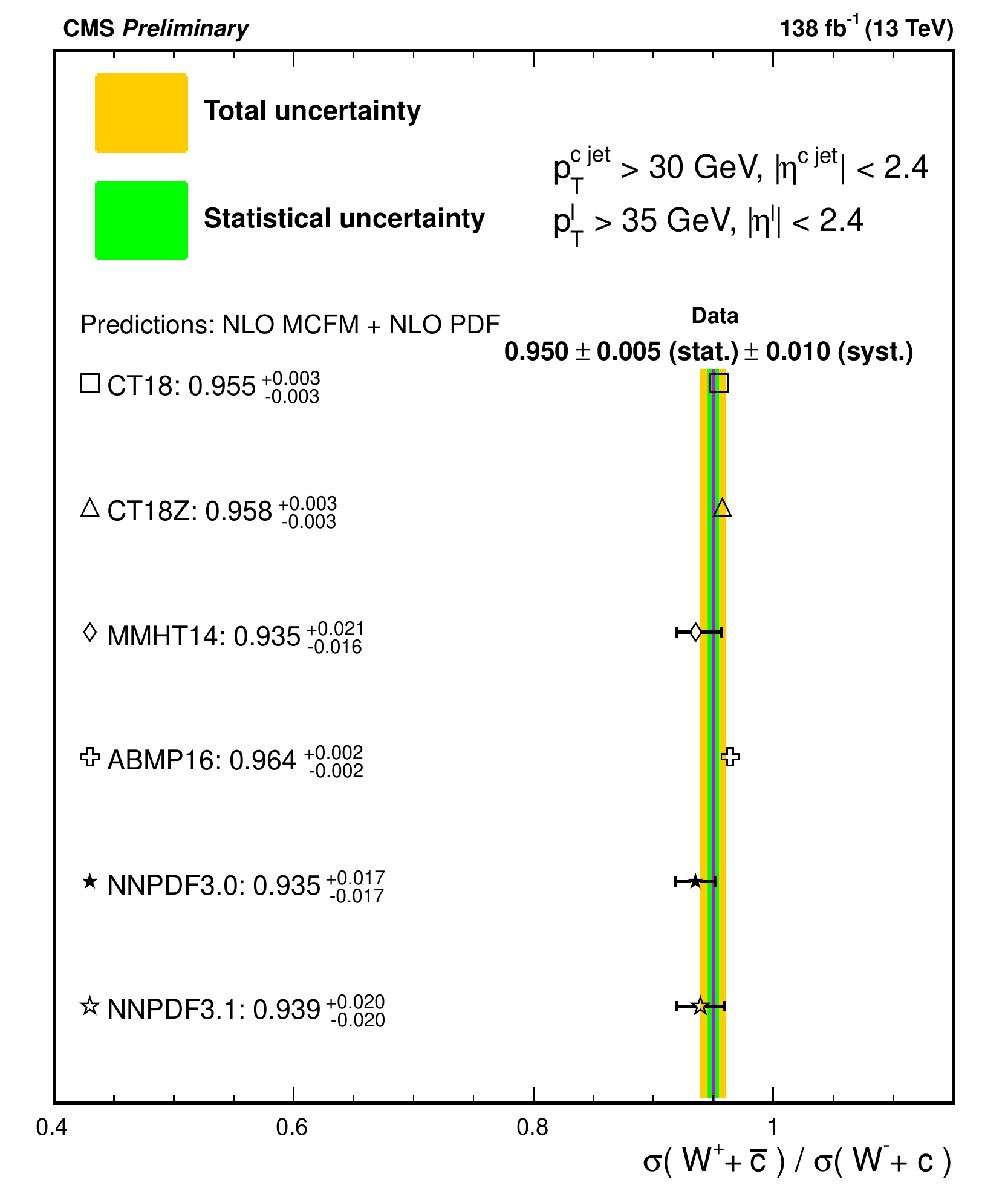}\hfill\vspace{0pt}
	\caption{Comparison of the theoretical predictions for the sum (left) and ratio (right) of $\sigma(\text{W}^+\bar{\text{c}})$ and $\sigma(\text{W}^-\text{c})$ computed with MCFM and several sets of PDFs with the CMS measurements at 13\TeV~\cite{CMS:2022bjk}. 
	}\label{fig3}
\end{figure}

Proton content in heavy flavour quarks can be probed from measurements at the LHC of the production cross section of a W or Z boson in association with a heavy flavour jet. In particular, the strange PDF can be constrained from the W$+$c production measurement. Beyond PDFs, W$+$c production allows test of pQCD and as a background to other analyses, controlling its modelling is essential. The cross section has been measured at $\sqrt{s}=13(8)\TeV$ from the $e\nu$ and $\mu\nu$ W decay channels in the fiducial region, defined by $\eta(\ell) < 2.4(2.1)$, $\pt(\ell)>35(30)\GeV$, and a c-quark with $\pt>25\GeV, |\eta|<2.5$ in jets defined as in the previously described measurements. The c-jet tagging is performed by either requiring a muon in the jet or a secondary vertex leading to two event categories which are combined. The results on the $\sqrt{s}=13\TeV$ total cross section and $\text{W}^+\bar{\text{c}}$ to $\text{W}^-\text{c}$ ratio in the fiducial region is compared with NLO predictions obtained with MCFM~\cite{Campbell:2019dru} using different PDFs in Fig~\ref{fig3}. More results, for $\sqrt{s}=8\TeV$ and the differential cross section as function of the lepton \pt and $\eta$ can be found in Refs.~\cite{CMS:2021oxn,CMS:2022bjk}. The bottom quark content of the proton can be constrained from the measurement of the production of a Z boson in association with a b quark. The measurement is also important to test the cross section calculation which can be done in the four or five flavour schemes with different approximation. The results of the CMS measurement at $\sqrt{s}=13\TeV$ which was presented can be found in Ref.~\cite{CMS:2021pcj}.

\section{Z boson invisible width measurements}

The production at the LHC of a Z boson in association with a jet has been exploited to measure the Z boson invisible width. The width is extracted from the ratio of the cross sections of the Z($\rightarrow\nu\bar{\nu}$)+jets and Z($\rightarrow\ell^+\ell^-$)+jets, $\ell = \mu, \text{e}$ production at the LHC and the LEP measurement~\cite{10.1093/ptep/ptaa104} of $Z\rightarrow(\ell\ell)$ width. Data driven methods are used to estimate the background to subtract and the cross sections are obtained from a fit of the dilepton \pt, or of the hadronic recoil (for Z$\rightarrow\nu\bar{\nu}$), distributions: $\Gamma(\text{Z}\rightarrow\nu\bar{\nu}) = \sigma(\text{Z}(\rightarrow\nu\bar{\nu})+\text{jets}) / \sigma (\text{Z}(\rightarrow\ell\ell)+\text{jets}) \cdot \Gamma(\text{Z}\rightarrow\ell\ell)$

The invisible width has been measured as $\Gamma_{\text{inv}} = 523 \pm 3 (\text{stat}) \pm 16 (\text{syst})\MeV$, with an accuracy similar to the LEP combined measurement~\cite{ALEPH:2005ab,10.1093/ptep/ptaa104}. The different contributions to the systematic uncertainties are shown in Table~\ref{tab1}. The result is compatible with the Standard model prediction, $\Gamma_{\text{inv}} = 504 \pm 0.04\MeV$. It is compared with the LEP measurement in Fig.~\ref{fig4}. More details on this measurement can be found in Ref.~\cite{CMS:2022ett}.
\begin{table}[htpb]
\begin{minipage}[b]{0.49\textwidth}
  \centering
  \includegraphics[width=0.95\textwidth]{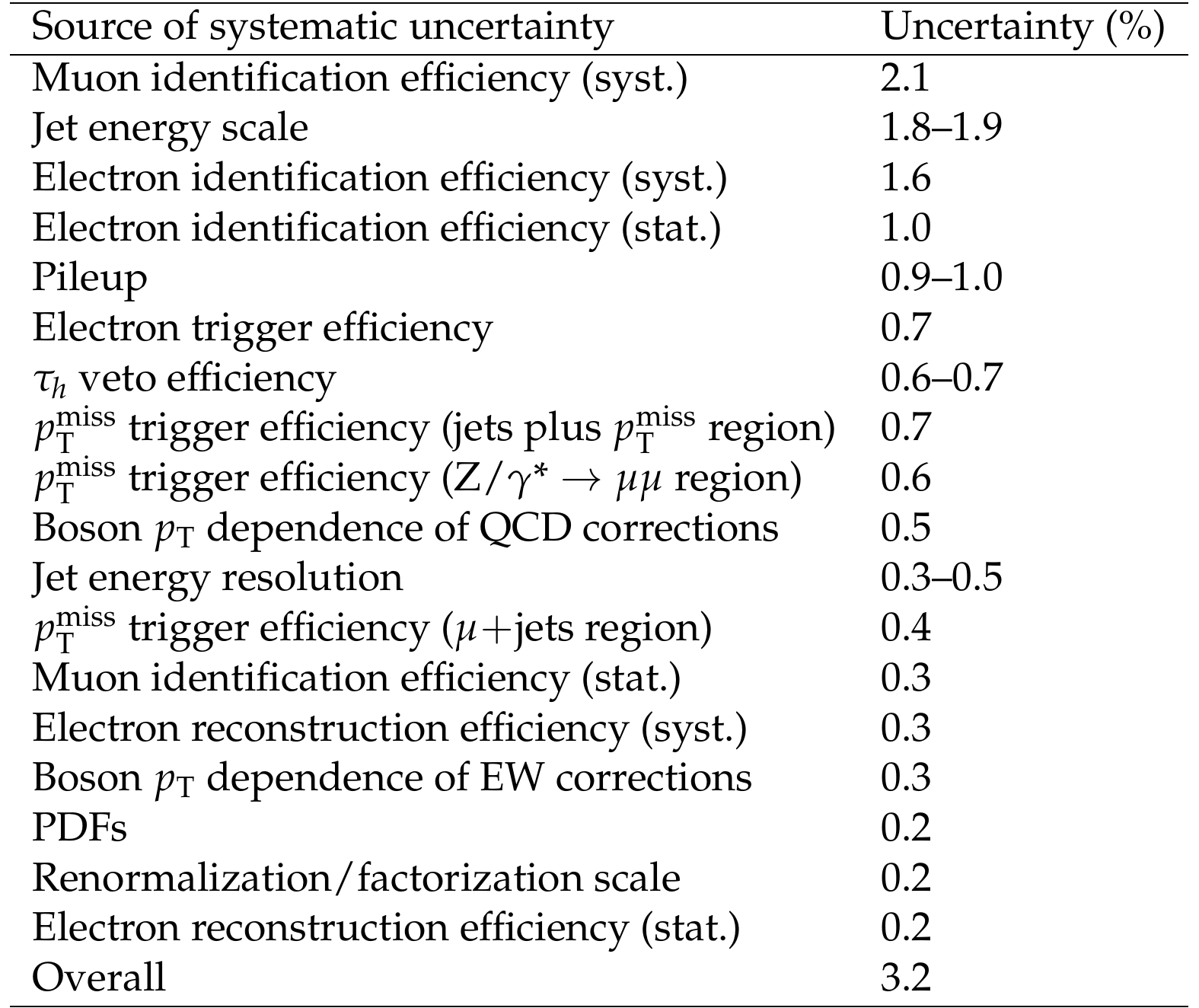}
  \captionof{table}{Relative uncertainties (in \%) on the final measurement from different sources~\cite{CMS:2022ett}.}\label{tab1}
\end{minipage}\hfill
\begin{minipage}[b]{0.49\textwidth}
\includegraphics[width=\textwidth]{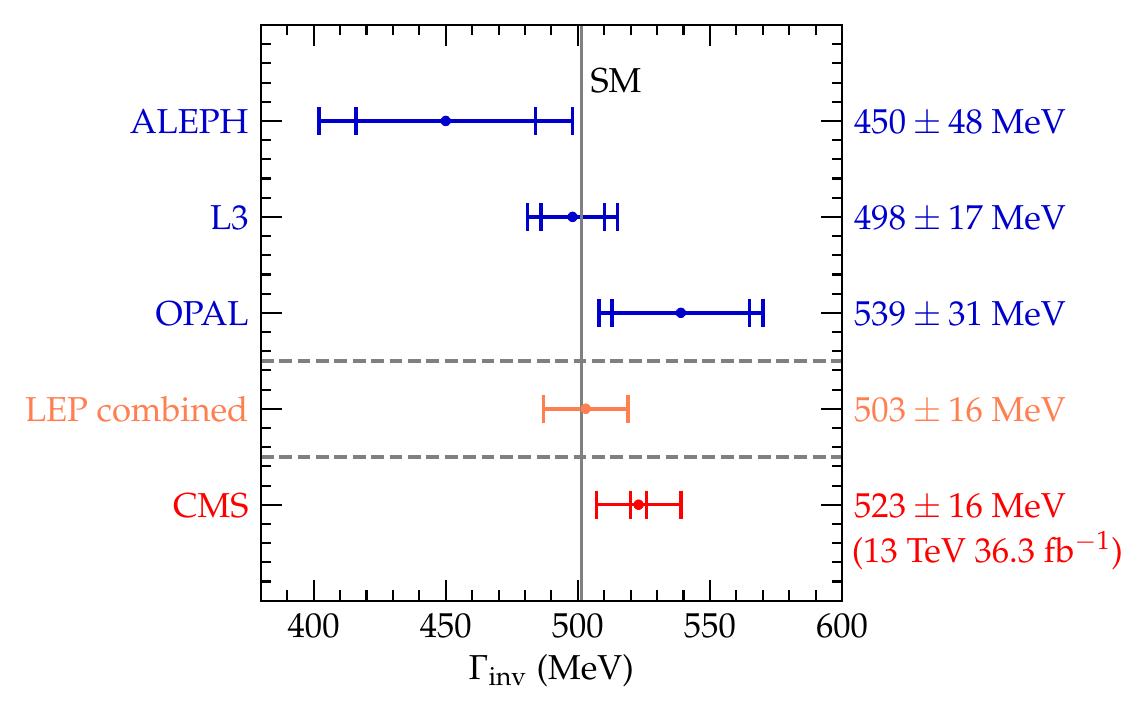}
\captionof{figure}{Direct measurements of the Z invisible width by the LEP experiments and the result from the CMS experiment presented here~\cite{CMS:2022ett}. Also shown is the prediction from the SM. The inner (outer) error bars show the statistical (total) uncertainty.}\label{fig4}
\end{minipage}
\end{table}

\section{Summary}

The latest CMS results on the production at the LHC of an electroweak boson in association with jets have been presented. The Z/$\gamma^*$ production with jets has been measured over a wide range of dilepton masses above 50\GeV. The full LHC run 2 data set has been exploited to measure the W$+$c and Z$+$b($\bar{\text{b}}$) production. Finally, a measurement of the invisible Z boson width has been obtained with the data acquired in 2015 and 2016 reaching a precision similar to the combined LEP result. These results complement a large legacy of measurements performed at the 7, 8, and 13$\,$TeV centre-of-mass energy.

\bibliography{pgras-ichep2022}
\bibliographystyle{JHEP}

\end{document}